\documentclass[sigconf]{acmart}
\AtBeginDocument{%
  }

\setcopyright{acmlicensed}
\copyrightyear{2018}
\acmYear{2018}
\acmDOI{XXXXXXX.XXXXXXX}

\acmConference[Conference acronym 'XX]{Make sure to enter the correct
  conference title from your rights confirmation emai}{June 03--05,
  2018}{Woodstock, NY}
\acmISBN{978-1-4503-XXXX-X/18/06}

\usepackage{tabularx}
\usepackage{graphicx}
\usepackage{subcaption}
\usepackage{chato-notes}

\theoremstyle{definition}
\newtheorem{definition}{Definition}[section]

\begin{document}

\title{Understanding and Addressing Gender Bias in Expert Finding Task}

\author{Maddalena Amendola}
\email{maddalena.amendola@phd.unipi.it}
\affiliation{%
  \institution{University of Pisa}
  \city{Pisa}
  \country{Italy}
}

\author{Carlos Castillo}
\email{chato@icrea.cat}
\affiliation{%
  \institution{Universitat Pompeu Fabra}
  \city{Barcelona}
  \country{Spain}
}

\author{Andrea Passarella}
\email{andrea.passarella@iit.cnr.it}
\affiliation{%
  \institution{IIT-CNR}
  \city{Pisa}
  \country{Italy}
}

\author{Raffaele Perego}
\email{raffaele.perego@isti.cnr.it}
\affiliation{%
  \institution{ISTI-CNR}
  \city{Pisa}
  \country{Italy}
}

\renewcommand{\shortauthors}{Trovato et al.}

\begin{abstract}
The Expert Finding (EF) task is critical in community Question\&Answer (CQ\&A) platforms, significantly enhancing user engagement by improving answer quality and reducing response times. However, biases, especially gender biases, have been identified in these platforms. This study investigates gender bias in state-of-the-art EF models and explores methods to mitigate it.
Utilizing a comprehensive dataset from StackOverflow, the largest community in the StackExchange network, we conduct extensive experiments to analyze how EF models' candidate identification processes influence gender representation. Our findings reveal that models relying on reputation metrics and activity levels disproportionately favor male users, who are more active on the platform. This bias results in the underrepresentation of female experts in the ranking process.
We propose adjustments to EF models that incorporate a more balanced preprocessing strategy and leverage content-based and social network-based information, with the aim to provide a fairer representation of genders among identified experts. Our analysis shows that integrating these methods can significantly enhance gender balance without compromising model accuracy.
To the best of our knowledge, this study is the first to focus on detecting and mitigating gender bias in EF methods.
\end{abstract}

\received{20 February 2007}
\received[revised]{12 March 2009}
\received[accepted]{5 June 2009}

\maketitle

\section{Introduction}
In recent years, community Question\&Answering (CQ\&A) platforms have emerged as very important tools for knowledge exchange and problem-solving. These collaborative forums allow users to ask questions and receive answers from experts within the community, promoting an atmosphere of shared learning and support. A key component of these platforms is the Expert Finding (EF) task, which aims to identify and rank users most likely to provide high-quality answers to new questions. Effective EF not only boosts user engagement but also improves answer quality and reduces response times, thereby enhancing the overall utility of the platform.
Despite significant advances in the accuracy of EF models, CQ\&A platforms face critical issues related to biases against minority groups. On StackOverflow (SO), one of the largest CQ\&A platforms, women make up less than 10\% of the user base. In addition, even more pronounced bias can manifest in the selection of experts to answer queries and biased ranking of answerers. Such biases undermine the fairness of the platform and discourage participation from underrepresented groups, further sustaining the disparity.

This work investigates gender bias in state-of-the-art EF models and identifies mechanisms to mitigate it. Our extensive experiments utilize a large-scale dataset from StackOverflow, the largest scientific community within the StackExchange network, where significant evidence of gender bias has been documented in previous studies \cite{wang2018understanding,tsay2014let,ford2016paradise,hanlon2018stack,maftouni2022thank,brooke2019condescending,scheltens2022representation,may2019gender,brooke2021trouble,dev2019quantifying,morgan2017programming,ford2017someone}. We examine how the selection of expert users affects gender balance and propose adjustments to existing models to ensure fairer outcomes.
Our findings indicate that the preprocessing strategies used in EF tasks, i.e., the process to identify candidate sets of experts, which are then ranked to obtain a final prioritised list, significantly influence the gender representation among identified experts. For instance, models that rely on reputation metrics or the number of UpVotes tend to favor men, who are generally more active on the platform, though not necessarily more competent. This results in the underrepresentation of women, placing them lower in the rankings and amplifying existing gender biases.
Then, we conduct an in-depth analysis of the model that has proven to best balance fairness and accuracy among the state-of-the-art EF models considered. By adapting the preprocessing phase, this model remains closer to the expected percentage of women in the dataset, stabilizing gender representation.
Additionally, we emphasize the importance of combining content-based and social network-based information to avoid biases inherent in each method. Models that rely solely on content may select more men, while those based solely on social information may overrepresent women due to peer-parity dynamics.


\section{Related Work}
Our research is related to studies investigating the presence of biases in CQ\&A communities in general and in SO in particular. 

\vspace{1mm}
\noindent \textbf{Biases in CQ\&A  platforms.}
Online communities offer benefits not only to content consumers but also to content creators. These platforms allow contributors to showcase their skills and expertise \cite{perkel2008peer,xu2020makes}. Research has shown that men and women participate and contribute differently in rates and types of involvement. These differences vary across different types of communities and evolve over time \cite{dubois2022gender}. Additionally, in communities with a gender imbalance, the content tends to reflect the interests of the dominant group \cite{reagle2011gender,das2019gendered}.
Even in communities with a relatively balanced gender ratio, such as Graphic Design communities, differences in participation and bias mechanisms have been identified. Dubois et al. \cite{dubois2020gender} found that on Stack Exchange, women tend to answer more opinion-based questions, receive less recognition, and express greater confidence in their language than men. In contrast, Quora shows fewer gender differences, with women more likely to respond to older questions. 
The same authors \cite{dubois2022towards} suggest that community presence information can humanize CQ\&A interactions, increasing user empathy and trust in the content. Furthermore, homophily and heterophily can play complementary roles in promoting an inclusive environment, particularly among women.
Open Source Software (OSS) communities similarly rely on voluntary contributions. Singh \cite{singh2019women} highlights the discrimination and isolation experienced by women in these communities, emphasizing the importance of "peer parity," where the presence of peers helps women feel included. Similarly, Wachs et al. \cite{wachs2017men} found that women in Dribbble have more clustered and gender-homophilous following relations, resulting in smaller and more closely knit social networks. 
Sun et al. \cite{sun2020male} examined a Python community and found that male users tend to provide informational help. In contrast, female users prefer participating in topics related to making friends and advertising. Furthermore, female users express positive emotions more frequently, and their activity is more susceptible to emotional orientation. Liu et al. \cite{liu2018research} found similar differences in online health communities, where women users are more inclined to seek emotional support, expressing more anxiety and sadness. Regarding networks, male users were more centered and influential than women.
Terrell et al. \cite{terrell2017gender} found that in open-source communities, women's pull requests are accepted more often than men's, but only when their gender is not identifiable.
Lin et al. \cite{lin2020examining} argue that gender influences how individuals share information and make decisions. They found that women prioritize social ties and commitment more than men when forming their attitudes.
Gender disparity has also been observed on online learning platforms. Wang et al. \cite{wang2023mind} found that questions from female learners are less likely to receive responses, attributing this to a male-driven gender homophily mechanism. 
Biases are also evident in Wikidata, with less than 22\% of items representing people being about women. This gender distribution is skewed towards men, mainly due to Wikidata editors oversampling male-dominated professions \cite{zhang2021quantifying}. 
In OpenStreetMap, women are dramatically underrepresented, contributing different types of content and focusing on different places compared to men \cite{das2019gendered}.
Finally, Hannak et al. \cite{hannak2017bias} show that in online freelancing marketplaces, perceived gender and race significantly influence worker evaluations, potentially harming employment opportunities, with women receiving fewer reviews.

\vspace{1mm}
\noindent \textbf{Biases in StackOverflow.}
CQ\&A platforms, such as SO, are designed to promote knowledge sharing and user engagement. These platforms are often perceived as meritocratic and free of gender barriers due to their openness and transparency \cite{wang2018understanding,tsay2014let}. However, several studies reveal significant gender biases impacting user experiences and outcomes on these platforms.
Research indicates that SO tends to be a male-dominated environment \cite{wang2018understanding, tsay2014let}. For instance, Ford et al. \cite{ford2016paradise} conducted semi-structured interviews and surveys, uncovering that women face several barriers more than men. These barriers include doubts about their expertise and feeling overwhelmed by the competitive environment. Jay Hanlon, vice president of community growth at SO, acknowledged the presence of race and gender biases, stating that many perceive SO as hostile or elitist, particularly newer coders, women, people of color, and other marginalized groups \cite{hanlon2018stack}.
SO has frequently been criticized for being a harsh and unfriendly environment \cite{maftouni2022thank}. In a 2019 survey, SO asked nearly 80,000 users what aspects of the platform they would most like to change, revealing gender-based differences in perceptions. Men associated the platform with terms like "official", "complex", and "algorithm", while women described it as "condescending," "rude," and "assholes" \cite{brooke2019condescending}.
The gender imbalance on SO is evident, with women consistently making up less than 10\% of participants. Common challenges women face include fear of negative feedback, lower confidence in their programming skills, an unwelcoming environment, inappropriate language, the competitive nature of the platform, and lack of peer support \cite{scheltens2022representation}.
Furthermore, evidence suggests that men benefit more from the current reputation system, which is biased against women. This disparity arises from gender differences in participation: women are more likely to ask questions, while men are more likely to provide answers and cast votes. The system favors answering questions, disadvantaging women due to their higher tendency to ask questions \cite{wang2018understanding, may2019gender}. Studies show that the average woman has roughly half the reputation points of the average man \cite{may2019gender, brooke2021trouble}. Male users tend to receive higher scores for their answers, suggesting that biases in scoring contribute to gender disparities in reputation \cite{brooke2021trouble}.
Additionally, votes on questions and answers are influenced by reputation bias, where users are more likely to vote positively on content from users with higher reputations, regardless of content quality \cite{dev2019quantifying}.
Finally, Ford et al. \cite{ford2017someone} suggest that the presence of more women in a thread creates a supportive environment, boosting female participation and fostering a sense of belonging and mentorship. Women become more active after engaging in peer parity posts, defined as interactions where individuals can identify with at least one other peer \cite{morgan2017programming}. Brooke \cite{brooke2021trouble} concludes that SO users tend to interact with others of the same or similar gender, indicating a gender-based organization in user interactions. 

In conclusion, while SO aims to be an open and meritocratic platform, significant gender biases persist, affecting user experiences and outcomes. Addressing these biases is crucial to creating more inclusive and supportive communities for all users.

\section{Background on EF and Gender Bias}
In this section, we explore the primary approaches to addressing the EF task, relying on various information sources, which form the core of our analysis and mitigation actions. We then discuss the two prevalent methodologies commonly used to evaluate EF solutions, offering a new perspective on this critical aspect.
\vspace{-6pt}
\subsection{Expert Finding Methods}
\label{sec:relatedEF}
The main sources of information for addressing the EF task come from the content of historical questions and answers and the interactions among users that can be modelled using hand-crafted features or network-based models.

\vspace{1mm}
\noindent 
\textbf{Text-based.}
Various methods have been proposed to address the EF task by leveraging similarities between current and previously answered questions. Liang et al. \cite{DBLP:conf/www/Liang19} propose an unsupervised generative adversarial network to measure similarities between word representations and experts.
Similarly, Dehghan et al. \cite{DBLP:journals/tkdd/DehghanA19} introduce a model that clusters question terms based on semantic similarity and co-occurrence. The same authors in \cite{DBLP:journals/ipm/DehghanBA19} further model user expertise by considering the tree structure of different domains, tags, and the temporal dimension of user answering behavior. In a complementary approach, Zhang et al. \cite{DBLP:conf/wsdm/ZhangCZCXLC20} tried to capture temporal dynamics by proposing models with multi-shift and multi-resolution settings.
Fu et al. \cite{fu2020recurrent} introduce a Recurrent Memory Reasoning Network that uses reasoning memory cells with attention mechanisms to focus on different aspects of the question. Additionally, Peng et al. \cite{peng2022towards, peng2022towards_comprehensive, liu2022expertbert, peng2023contrastive} utilize attention mechanisms in multi-view, multi-grained, semi-supervised pre-trained, and pre-trained with personalized fine-tuning models for expert finding.
Finally, Qian et al. \cite{qian2022multi} propose a Multi-Hop Interactive Attention-based Classification Network with attention mechanisms to capture latent interactions among question subjects and bodies.

\vspace{1mm}
\noindent \textbf{Feature-based.}
The second group of methods in expert finding relies on hand-crafted features to model community members' expertise. This approach incorporates various aspects of a user's activity and behavior. Roy et al. \cite{roy2018finding} introduce a scoring function that captures different dimensions of expertise, including profile tags, accepted answers, and recent activity. Mumtaz et al. \cite{mumtaz2019expert2vec} extend this foundation by proposing a framework that integrates activity, community, and time-aware features. Similarly, Fu et al. \cite{fu2019tracking} and Kundu et al. \cite{kundu2019finding} emphasize the temporal aspect, tracking the evolution of user roles.
To further refine the evaluation of expertise, Tondulkar et al. \cite{tondulkar2018get} propose features favoring experts who provide high-quality answers to complex questions, incorporating LtR methods. Differently, Faisal et al. \cite{faisal2019expert} introduced a model based on an adaptation of the bibliometric g-index, emphasizing answer quality and consistency. 
Fu et al. \cite{fu2020user} highlight the importance of the intimacy between the asker and answerer. This is complemented by Tan et al. \cite{tang2020hierarchical}, which combines factorization machines with hierarchical attention mechanisms, capturing user-expert interactions and modeling the importance of different features.
Finally, Roy et al. \cite{roy2024early} diverge from traditional approaches by aiming to predict promising expert users at an early stage, rather than focusing solely on the most senior platform users.

\vspace{1mm}
\noindent \textbf{Network-based.}
Network-based methods integrate network interactions and relationship information. Kundu et al. \cite{kundu2019formulation} introduce a framework combining a text-based component for estimating expertise with a Competition Based Expertise Network (CBEN) \cite{aslay2013competition} that uses link analysis techniques. The same authors further developed the framework to include intra-profile and inter-profile preferences \cite{kundu2020preference}, and later into a topic-sensitive hybrid expertise retrieval system (TSHER) integrating knowledge, reputation, and authority estimators \cite{kundu2021topic}.
Le et al. \cite{le2018retrieving} measure similarity between answerer and asker using social network techniques, while Sun et al. \cite{DBLP:conf/icwsm/SunMR018} build a competition graph representing hierarchical relationships among users and questions. This hierarchical concept is further explored \cite{DBLP:conf/aaai/SunBBLS019}.
To address data sparsity, Sang et al. \cite{sang2019multi} propose a Multi-modal Multi-view Semantic Embedding (MMSE) framework, integrating local and global views with social structure information. Ghasemi et al. \cite{ghasemi2021user} focus on user embedding, proposing a joint model for text and node similarity.
Li et al. \cite{li2019personalized} model a CQA platform as a Heterogeneous Information Network (HIN), using Metapath-based Embedding to rank experts based on a CNN scoring function. This approach is further developed by Qian et al. \cite{qian2022heterogeneous}. Liu et al. \cite{liu2022high} and Krishna et al. \cite{krishna2023temporal} address interest drift, modeling expert assessment based on relevance and depth within specific domains, and proposing a graph diffusion model to learn users' expertise semantically and temporally. Costa et al. \cite{costa2023here, costa2023ask} study temporally-discounted, tag-based models for EF tasks. Finally, in \cite{amendola2024towards}, authors model the CQ\&A platform using a Multi-Layer Graph (MLG) where each layer identifies a main discussed topic. The selection of experts is performed through RW, and they adopt Learning-to-Rank (LtR) techniques to rank the experts.

\subsection{Evaluating Expert Finding Methods}
\label{sec:EFeval}
Given a question posted by a user of the online community, the EF task aims to \textit{compute a short, ranked list of community members, i.e., experts, that are likely to provide an accurate answer to the question.}
The EF task can be interpreted as an item recommendation task. 
Given the large pool of items (users) to retrieve from, recent studies have adopted a target set approach, which speeds up metric calculation by evaluating models on a smaller subset of relevant items alongside a defined number of non-relevant items \cite{krichene2020sampled, dallmann2021case}. 
Recent research has highlighted significant concerns regarding sampled metric strategies. Krichene et al. \cite{krichene2020sampled} found that sampled metrics differ substantially from their exact versions, even in terms of relative statements, like comparing recommender A to recommender B. They noted that smaller sample sizes show fewer differences between metrics. Canamares et al. \cite{canamares2020target} discovered that comparative evaluations using reduced target sets often yield conflicting outcomes compared to evaluations using large target sets. Finally, Dallmann et al. \cite{dallmann2021case} studied two prevalent sampling methods: popularity sampling and uniform random sampling. They found both methods can produce rankings inconsistent with the models' full rankings and vary greatly across different sample sizes. 

In the EF context, identifying during first stage expert users based on criteria (i.e. minimum number of high-quality answers) simplifies the task and reduces complexity. However, this can exacerbate the cold-start problem, excluding new users who might provide valuable answers. Nonetheless, this approach speeds up metric computation and improves system accuracy.
So far, different strategies have been adopted to define a subset of users representing the experts: 10\% of the most active users \cite{kundu2021topic,qian2022heterogeneous, tang2020hierarchical,peng2022towards,peng2022towards_comprehensive,liu2022expertbert,peng2023contrastive,roy2024early,li2019personalized,kundu2019formulation,liu2022efficient}, all users with a minimum number of accepted answers \cite{kundu2021time, krishna2023temporal, qian2022multi, sun2018qdee, sun2019atp, fu2020recurrent, fu2020user, kundu2020preference, fu2019tracking}, or a two-stage approach that considers the acceptance ratio \cite{nobari2020quality, fallahnejad2022attention, kasela2023se, dargahi2017skill, mumtaz2019expert2vec}.
There is no established method to determine whether a user qualifies as an expert. However, for a fair comparison of state-of-the-art models for the EF task, it's essential to distinguish studies by their sampling approach during the ranking phase and ensure that the models are compared under the same assumptions. Specifically, we recognize two different methods:
\begin{itemize}
    \item \textbf{Experts Ranking}: for a new question, this group of works \cite{nobari2020quality,kundu2021topic,fallahnejad2022attention,liu2022high,costa2023here,kundu2021time,costa2023ask,krishna2023temporal,qian2022multi,roy2024early,sun2019atp,fu2019tracking,fu2020recurrent,mumtaz2019expert2vec,kundu2019formulation,sun2018qdee,fu2020user,kundu2020preference} ranks all the users labelled as expert users in the first stage. 
    \item \textbf{Experts Subsample Ranking}: given the new questions along with the information of all the users that answered the question (ground truth), the studies belonging to this group \cite{sang2019multi,ghasemi2021user,qian2022heterogeneous,tang2020hierarchical,sorkhani2022feature,peng2022towards,peng2022towards_comprehensive,liu2022expertbert,peng2023contrastive,li2019personalized,DBLP:conf/wsdm/ZhangCZCXLC20,liu2022efficient} generally rank only the users of the ground truth, or a small set of users (i.e. 20) that always include the ground truth plus other users usually randomly chosen from the 10\% of the most active ones. This approach resorts to sample strategies that are unreliable for evaluating a recommender system \cite{krichene2020sampled,canamares2020target,dallmann2021case}.
\end{itemize}

\subsection{Computing Gender and Limitations}

SO does not require users to state their gender identity \cite{wang2018understanding}. Various studies \cite{wang2018understanding,scheltens2022representation,may2019gender,brooke2021trouble,ford2017someone} have utilized genderComputer, a Python tool developed using data from SO. This tool infers gender using lookup tables and heuristics. It takes a (name, country) tuple as input and returns one of “female,” “male,” or “unisex” (when no gender can be inferred). The resolution algorithm starts with identifying the first and last names, then continues with gender detection based on gender-specific last name forms, country-specific lookup tables, cross-country lookup, and diminutive resolution \cite{vasilescu2012gender}. 
We acknowledge several limitations and drawbacks in our approach to inferring gender. First, we simplify gender as binary. Computational studies should consider how individuals perform their gender identity within specific social settings and how these performances are influenced by broader social structures and expectations \cite{brooke2019condescending}. 
We argue that this is a fundamental limitation and that further research is needed to include the perspectives of non-binary users. 
Second, as noted by \cite{may2019gender}, there are limitations related to the geographic component of our inference: a minority of users include location data, and a given location may not reflect a user’s origin. Moreover, anonymity can impact behavior \cite{robertson2018estimates}, and users can decide to hide their real identity through anonymity to feel more independent (known as \emph{gender swapping} \cite{bruckman1996gender,szell2013women}). In this context, women may pose as men if they feel that they will be taken more seriously or to avoid biases. Despite these limitations, we rely on the fact that genderComputer is currently the most widely used tool for computing gender when analyzing gender biases on SO.

\section{Analysing and measuring the bias in the SO dataset}

This study aims to evaluate the state-of-the-art models for the EF task on SO platform, specifically focusing on gender bias. By analyzing these models for gender bias, we aim to provide useful recommendations for improving their design. The implementation is publicly available\footnote{\url{https://bit.ly/genderbias_EF}} for reproducibility. 
The first step in this process is to select the relevant data and statistically test for gender differences.

\subsection{The StackOverflow dataset}
We conducted experiments using a large-scale dataset from SO, the largest community in the Stack Exchange network. In this community, users post questions by specifying a title, body, and tags to identify the topics better. Other users provide various answers to these questions, which can receive UpVotes or DownVotes. Most importantly, one answer is accepted as the best answer. The more UpVotes a user receives, and the more their answers are accepted, the higher their Reputation score will be, reflecting their ability to provide high-quality answers.
The publicly available SO data dump\footnote{\url{https://archive.org/details/stackexchange}} contains over 24 million questions from community members, spanning from July 31, 2008, to the present. Due to the dataset's size, we selected data spanning six years, from January 2017 to December 2022, for our analysis. 
To better capture the evolving nature of the SO community, our analysis covers distinct subsets of the dataset. Specifically, we analyze data from pairs of consecutive years, groups of three consecutive years, and the entire six-year period. For each time frame considered, we temporally split the dataset into training (first 80\%) and testing (remaining 20\%) sets. For example, a period labelled "2017-2018" in tables or figures refers to data from January 2017 to December 2018.
Table \ref{tab:dataset} presents key statistics for various data segments, organized by the period in the \textit{Year} column. It includes the number of training questions that received accepted answers (\textit{Questions}), the number of users who provided at least one accepted answer (\textit{Answerers}) during each specified period, and for whom we can compute the gender. 
The \%Female column reports the percentage of answerers who are female. Finally, the table shows the percentage of answers and accepted answers provided by women (\textit{\%AnswersFemales} and \textit{\%AcceptedFemales}, respectively).

\begin{table}[]
\caption{Characteristics of the StackOverflow datasets used.}
\centering
\resizebox{\columnwidth}{!}{
\begin{tabular}{crrcccc}
\hline
Years & Questions & Answerers & \%Females & \begin{tabular}[c]{@{}c@{}}\%Answers\\ Females\end{tabular} & \begin{tabular}[c]{@{}c@{}}\%Accepted\\ Females\end{tabular} \\\hline
2017-2018 & 1,243,241 & 69,162 & 9.20\% & 8.45\% & 8.12\% \\
2019-2020 & 1,096,295 & 60,527 & 9.42\% & 9.05\% & 8.68\% \\
2021-2022 & 803,947 & 40,730 & 9.51\% & 9.61\% & 9.23\% \\
2017-2019 & 1,784,363 & 89,824 & 9.37\% & 8.58\% & 8.25\% \\
2020-2022 & 1,372,516 & 64,682 & 9.61\% & 9.60\% & 9.29\% \\
2017-2022 & 3,172,359 & 134,130 & 9.74\% & 8.94\% & 8.62\% \\\hline
\end{tabular}}
\label{tab:dataset}
\end{table}

\begin{table}[]
\caption{Welch's t-test with \(\alpha=0.01\) and $10,000$ permutations between male and female features' distributions.}
\centering
\resizebox{\columnwidth}{!}{
\begin{tabular}{c|cc|ccccc}
\hline
Years & Answers & Accepted & Reputation & UpVotes & DownVotes & Views \\ \hline
2017-2018 & 0.09 (M) & 0.03 (M) & \underline{0} (M) & \underline{0} (M) & 0.44 (F) & 0.15 (M) \\
2019-2020 & 0.66 (M) & 0.47 (M) & \underline{0} (M) & \underline{0} (M) & 0.65 (F) & \underline{0} (M) \\
2021-2022 & 0.99 (M) & 0.81 (M) & \underline{0} (M) & \underline{0} (M) & 0.61 (F) & 0.01 (M) \\
2017-2019 & 0.08 (M) & 0.04 (M) & \underline{0} (M) & \underline{0} (M) & 0.46 (F) & 0.15 (M) \\
2020-2022 & 0.95 (M) & 0.82 (M) & \underline{0} (M) & \underline{0} (M) & 0.60 (F) & \underline{0} (M) \\
2017-2022 & 0.19 (M) & 0.12 (M) & \underline{0} (M) & \underline{0} (M) & 0.67 (F) & 0.08 (M) \\ \hline
\end{tabular}}
\label{tab:features_statistics}
\end{table}

\subsection{Comparative Statistics}

\begin{figure}
    \centering
    \includegraphics[width=0.7\linewidth]{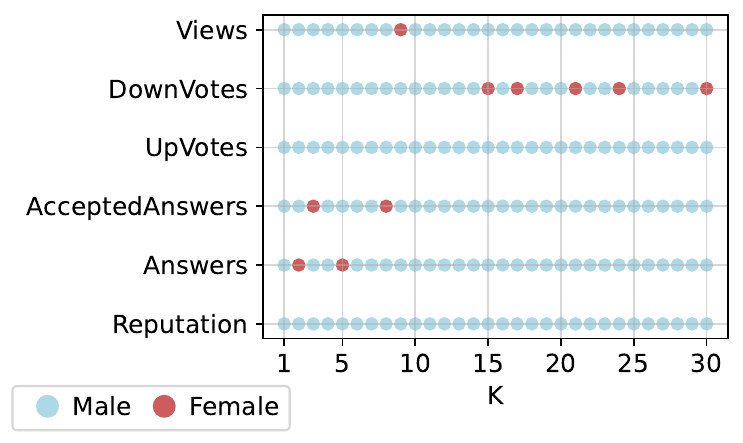}
    \caption{Experts sorting based on single features.}
    \label{fig:features_sorted}
\end{figure}

Different studies have already shown that male users are more active on the platform \cite{wang2018understanding,morgan2017programming,may2019gender,brooke2021trouble,scheltens2022representation}. This result is also reflected in Table \ref{tab:dataset}, which shows that the percentage of answers provided by females is, in most cases, lower than that of female users. This indicates lower participation from women compared to men. Notably, in 2021-2022, the percentage of answers from females is higher; in 2020-2022, the percentages are equal. Nevertheless, in both these periods (as well as in the other periods), the percentage of accepted answers from females is lower than their overall participation rate, indicating a lower tendency to accept answers from women.
Table \ref{tab:features_statistics} presents the results from Welch's t-test, with an $\alpha=0.01$ (standard threshold), assessing user features such as Reputation, number of Answers and Accepted Answers provided, number of UpVotes and DownVotes received, and number of profile Views across different dataset splits. The table details which gender showed higher average values for each feature and includes the p-values from the tests to indicate statistical significance. Underscored p-values showcase high statistical significance ($p-value<\alpha$).  We categorize Answers and Accepted Answers as \emph{participation} features since they are primarily related to users' engagement in the platform's primary task. The other features are denoted as \emph{visibility} features, as they relate more to the user's standing within the expert pool. Given the significant male majority in the data samples (over 90\%, as reported in Table \ref{tab:dataset}), we enhanced our analysis accuracy by employing 10,000 permutations for each test. 

Our analysis revealed that males consistently had higher averages across all metrics each year except for DownVotes, a negative reputation metric. Interestingly, participation metrics like Answers and Accepted Answers showed high p-values in half of the experiments, suggesting more similar distributions between genders and indicating a relatively balanced contribution level from both male and female users. In contrast, the lower p-values associated with visibility metrics in these experiments point to a significant disparity in how contributions are recognized, with male users frequently receiving more UpVotes, leading to higher Reputation scores.
The higher performance of males in visibility metrics may indicate a bias in how male and female contributions are received and valued, as demonstrated in \cite{brooke2021trouble}. The robustness of these findings, underscored by the extensive permutation testing, highlights persistent gender disparities on the platform over time. This discrepancy might influence personal perceptions of value and community influence, potentially discouraging female users if they perceive their contributions as less likely to be recognized or rewarded.
Additionally, our data indicated that females generally received more DownVotes than males. Despite the lack of statistical significance, the pattern of higher DownVotes for females merits attention as it could reflect subtler biases within community interactions. 

Figure \ref{fig:features_sorted} compares features among the top 30 users, differentiated by gender. Data points are colour-coded to represent gender, with red points indicating women and blue points indicating men. This colour coding allows for immediate visual comparison of gender-based performance across various metrics.
It is immediately apparent that men predominantly occupy the top positions across most metrics, particularly in UpVotes and Reputation. This trend suggests that men generally receive higher recognition and visibility on the platform. Regarding participation features such as Answers and Accepted Answers, women are present in the top positions but are largely absent beyond these initial ranks.
The discrepancy between the presence of women in the top ranks for Answers and Accepted Answers and their absence in the top ranks for Reputation, UpVotes, and Views underscores a significant gender difference in how contributions are evaluated and rewarded. Even though women actively participate and provide valuable answers, their contributions receive a different level of UpVotes, Reputation points, or visibility than their male counterparts.
Considering that women constitute only 9.30\% of the user base, the presence of women among the top 30 users suggests a relatively high level of activity or effectiveness within this minority group. However, the generally lower values for women across most metrics indicate potential systemic biases.

\begin{figure*}[h]
    \centering
    \begin{subfigure}[b]{0.24\textwidth}
        \centering
        \includegraphics[width=\textwidth]{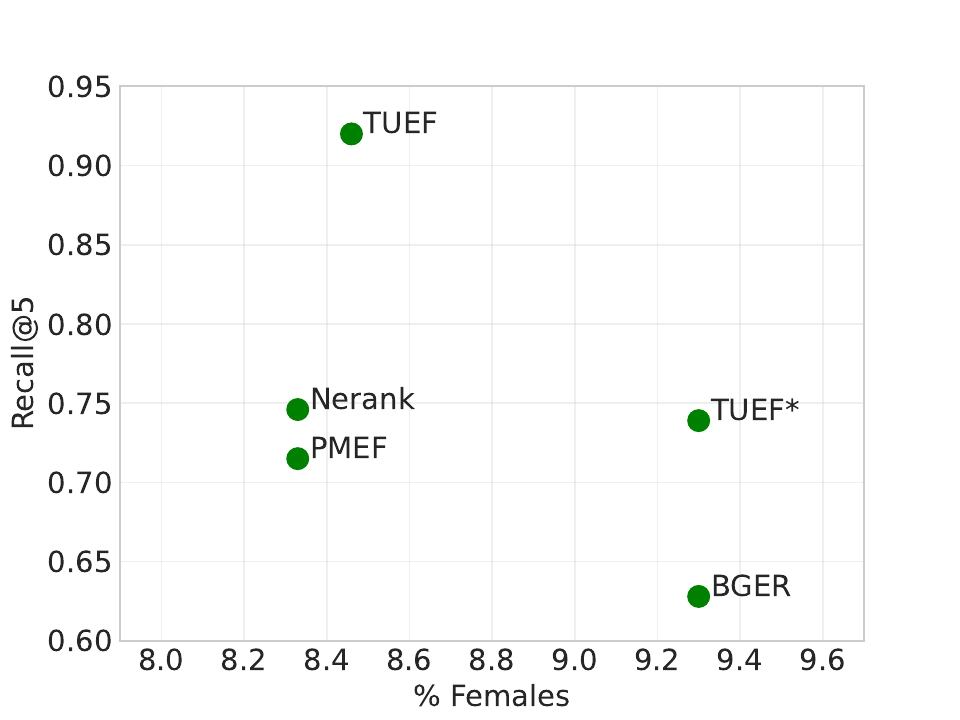}
        \caption{}
        \label{fig:recall_perc}
    \end{subfigure}
    \begin{subfigure}[b]{0.24\textwidth}
        \centering
        \includegraphics[width=\textwidth]{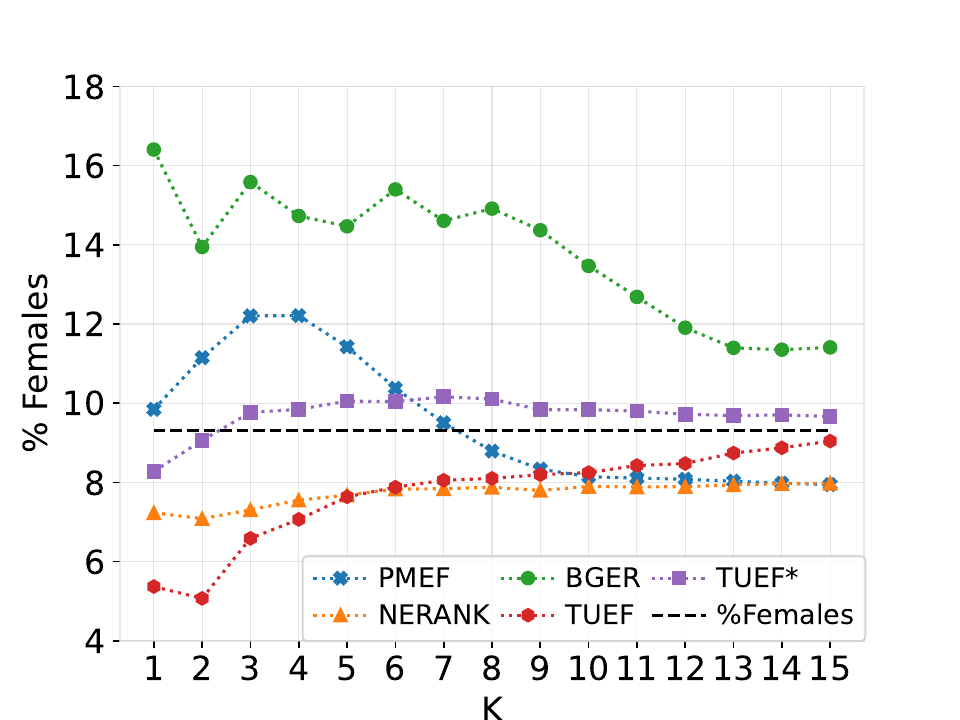}
        \caption{}
        \label{fig:perc_cutted_lists}
    \end{subfigure}
    \begin{subfigure}[b]{0.24\textwidth}
        \centering
        \includegraphics[width=\textwidth]{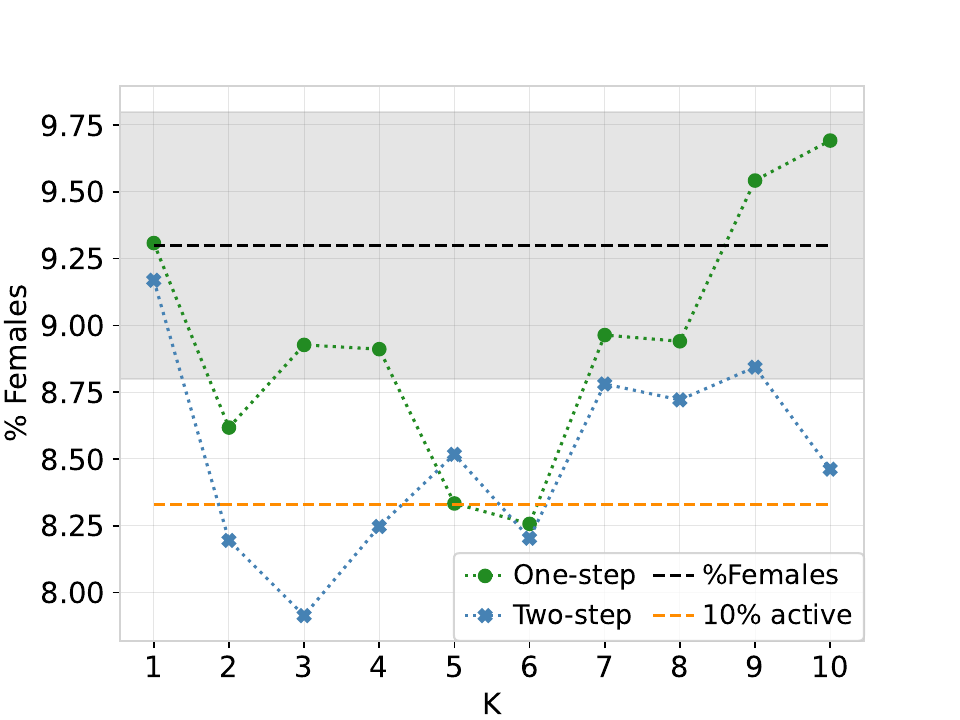}
        \caption{}
        \label{fig:perc_preprocess}
    \end{subfigure}
    \begin{subfigure}[b]{0.24\textwidth}
        \centering
        \includegraphics[width=\textwidth]{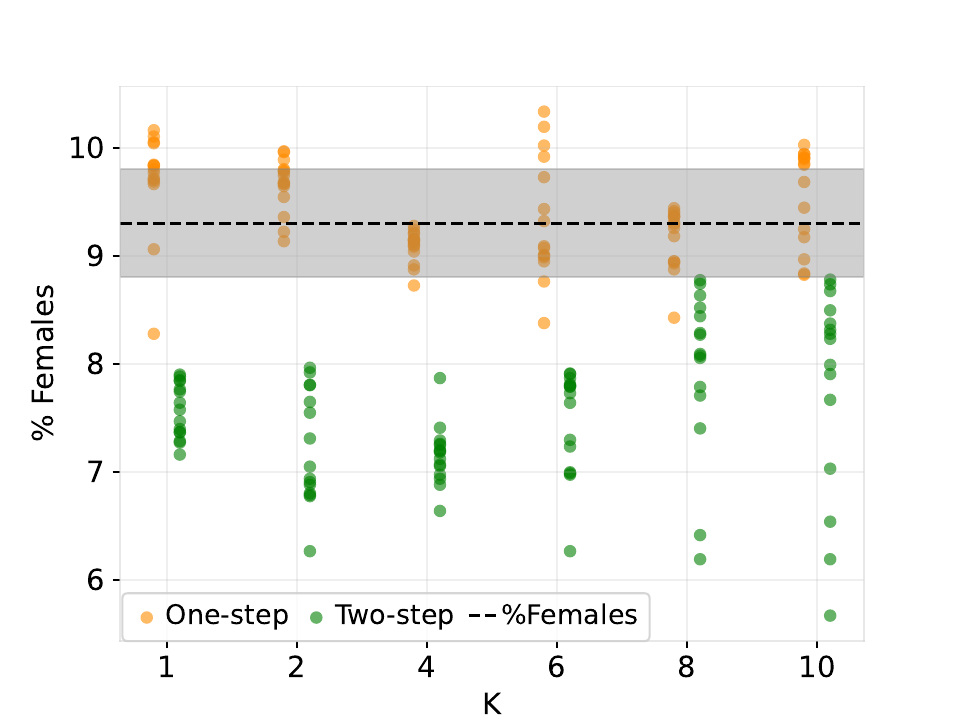}
        \caption{}
        \label{fig:tuef_perc_preprocess}
    \end{subfigure}
    \caption{Figure (a) compares the baselines considering Recall@5 (y-axis) and \% Females (x-axis); Figure (b) compares the baselines' ranking in terms of \% Females (y-axis) at different list cutoffs (K); Figure (c) compares the three preprocessing methods (varying the minimum number of accepted answers on the x-axis) with respect to the expected percentage of females; Figure (d) compares the \% Females (y-axis) for TUEF executed with Two-step and One-step preprocessing.}
    \label{fig:4inrow}
\end{figure*}

\section{Bias in Expert Finding methods}
\label{sec:bias_ef}
This section aims to assess state-of-the-art models for the EF task, focusing on evaluating their performance regarding gender bias. To this end, we consider the following definition:
\begin{definition}
Given a population with a specific gender distribution, we consider a model fair if its outcomes accurately reflect these gender proportions.
\end{definition}
In the context of EF models, we expect them to maintain the same percentage of men and women as present on the CQ\&A platform, thus avoiding the amplification of existing biases.
We have selected four models for this analysis:
\begin{itemize}
    \item \textbf{NeRank \cite{li2019personalized}:} jointly learns embeddings of question content, question raisers, and question answerers using a combination of a Heterogeneous Information Network embedding algorithm and a Long Short-Term Memory model. It then uses a convolutional scoring function to identify experts.
    \item \textbf{PMEF \cite{peng2022towards}:} utilizes a multi-view attentive matching mechanism consisting of three modules: a question encoder, an intra-view encoder, and an inter-view encoder to understand the relationships between experts and questions.
    \item \textbf{BGER \cite{krishna2023temporal}:} a graph diffusion-based expert recommendation model that learns users’ expertise in the context of both semantic and temporal information to capture their changing interests and activity levels over time.
    \item \textbf{TUEF \cite{amendola2024towards}:} leverages content and social information by defining a topic-based MLG that models users’ similarities in providing answers.
\end{itemize}

NeRank and PMEF fall under the \emph{Experts Subsample Ranking} category defined in Section \ref{sec:EFeval}, while BGER and TUEF are categorized in Section \ref{sec:EFeval} as \emph{Expert Ranking} methods. The selection of these methods stems from the need to perform a bias analysis on EF methods that provide good coverage of the variety of state-of-the-art solutions discussed in Section \ref{sec:relatedEF}. Additionally, the availability of the source code for these solutions ensures fair and reproducible experiments.
NeRank and PMEF are designed to rank a set of 20 users, always including the actual answerers of the question, with additional users randomly chosen from the top 10\% most active users (10\% preprocessing). In contrast, BGER and TUEF rank all users who meet specific criteria. BGER includes all users with a minimum number of accepted answers (one-step preprocessing), while TUEF goes a step further by also selecting those whose acceptance ratio—the ratio of accepted answers to total answers—exceeds the average (two-step preprocessing).
Below, we compare the four models across recall and gender balance dimensions. Then, we consider the tradeoff between gender balance and accuracy in the expert recommendation. Next, we analyze the potential gender bias introduced by the preprocessing strategies adopted in the considered EF solutions. Finally, we perform an in-depth gender bias analysis of the different steps of TUEF, the method showing superior performance in the previous analysis.
\vspace{-6pt}
\subsection{Baselines comparison and bias mitigation}
\label{sec:preprocessing}

This section aims to compare the state-of-the-art baselines, identify potential biases, and propose methods to mitigate them. Due to the high computational demands of the neural network-based methods used in this analysis, we limit our dataset to the last 30,000 questions from the 2017-2022 dataset. To ensure a fair comparison of the four strategies, we select questions from the test set that meet two criteria: (i) the best answerer is identified as an expert by TUEF using a two-step selection process, and (ii) the asker has posted at least two questions in the training set. The latter criterion is a requirement for the NeRank and PMEF algorithms, while the former ensures a fair comparison between the sampling strategies of NeRank, PMEF, and TUEF.

The final testing set comprises 1,341 questions. In the training set, we identified 9,560 users who provided at least one accepted answer. GenderComputer was able to determine the gender for 3,728 of these users, with 9.30\% identified as female. Notably, this percentage aligns with the overall gender ratio in the dataset, underscoring the significance of the obtained results.
It is also important to highlight that the four models employ three different preprocessing methods, as detailed in Section \ref{sec:bias_ef}, which is crucial for the scope of the following analysis.

We tested all the models using the code provided by the authors. Figure \ref{fig:recall_perc} compares the four EF models across two dimensions. The x-axis represents the percentage of women included after the preprocessing phase performed by the algorithms: BGER includes all the answerers who have provided at least one accepted answer, maintaining a women's percentage of 9.30\%; NeRank and PMEF select the top 10\% of the most active users, encompassing 516 answerers, of whom 8.33\% are women; TUEF selects 130 answerers (8.46\% women). The y-axis shows the models' performance in terms of Recall@5.
TUEF outperforms NeRank and PMEF in both dimensions, maintaining a slightly higher percentage of women among the selected experts and achieving a relative performance increase of over 23\%. However, TUEF surpasses BGER only in terms of recall, as its preprocessing results in a lower-than-expected percentage of women.
To address this, we modified TUEF by removing the expert selection phase and treating all answerers as experts, thereby preserving the women's percentage at 9.30\%, similar to BGER. This adjustment denoted as TUEF*, reduces performance, aligning it with NeRank. However, TUEF* emerges as the most effective model, as it matches BGER on the x-axis and exceeds it in terms of recall. The decline in performance is reasonable, given that the model expanded its focus from considering only 130 answerers to more than 3,500.

Figure \ref{fig:perc_cutted_lists} displays the percentages of women predicted by each model at various cutoff points in the ranked lists. For any given value of K, a data point indicates the average percentage of women in the top K positions across all predictions. The dashed black line represents the expected percentage of women in the dataset. 
PMEF starts with a higher representation of women for smaller values of K and shifts to underrepresent women as K increases. NeRank and TUEF consistently underrepresent women across all values of K. On the other hand, BGER consistently underrepresents men, thereby typically showing a higher percentage of women.
Finally, TUEF*, modified to include all answerers as experts, closely aligns with the dashed black line. This indicates that TUEF* effectively mirrors the gender distribution of the dataset, making it the most balanced model regarding gender representation and fairness.

These results, particularly the marked differences in fairness between TUEF and TUEF*, illustrate how changes in preprocessing can enhance performance.
TUEF employs a two-step preprocessing approach, while TUEF* adopts a simpler one-step method with a threshold set to 1. Ideally, an effective EF system should assess and rank all users on the platform. However, given the vast size of this community, which exceeds 24 million users, a preprocessing phase is required to narrow the focus to more experienced users.

Figure \ref{fig:perc_preprocess} displays the percentage of females identified during the three preprocessing phases, measured against varying thresholds for the minimum number of accepted answers. The horizontal dashed black line represents the expected percentage of women within the dataset. The horizontal dashed orange line shows the percentage of women resulting from preprocessing that selects users from the top 10\% of most active users (therefore representing NeRank and PMEF). The green and blue dots connected by dashed lines illustrate the percentage of women when one-step and two-step preprocessing models are applied (thus, representing BGER, TUEF* and TUEF), respectively, setting different thresholds for the minimum number of accepted responses.
The analysis reveals that the preprocessing model that includes the top 10\% of users consistently underrepresents women, maintaining a steady percentage of around 8.33\%, which is significantly lower than the expected percentage. The one-step preprocessing model generally performs better, maintaining a female percentage close to or within a range of $\pm5\%$ (grey zone) around the expected value, especially as the threshold of minimum accepted responses increases. The two-step preprocessing model varies more widely and generally performs less consistently in representing women than the one-step model, particularly at higher thresholds.

Figure \ref{fig:tuef_perc_preprocess} illustrates scatter plots showing the percentage of women at various cutoffs in the prediction lists for TUEF, using one-step (green) and two-step (orange) preprocessing with different minimum accepted answers thresholds. The dashed horizontal line marks the expected percentage of women in the dataset. The plots reveal that the one-step preprocessing consistently maintains the percentage of women closer to the expected line across all cutoff points (K), suggesting a more stable and fair representation. In contrast, the two-step preprocessing shows greater variation and generally lower percentages of women, indicating less consistency in maintaining gender balance.

\section{The TUEF framework}
\label{sec:TUEF}

The analysis presented so far shows that TUEF (and its variation TUEF*) provides the best tradeoff among the considered benchmarks regarding accuracy and fairness. Therefore, in the rest of the paper, we analyze more deeply the impact of the various features of TUEF concerning fairness. Before, we provide some details about TUEF to interpret the results presented hereafter better.

TUEF is a topic-oriented model that uses tags associated with questions to represent the CQ\&A platform as a Multi-Layer Graph, where each layer represents a main topic. In each layer, user nodes are represented by \emph{knowledge vectors} that indicate their expertise under a specific topic, and edges are formed based on the cosine similarity between pairs of nodes, modelling similarities in answering behaviours.
For a new question, TUEF focuses its search on the relevant layers. It uses two methods: the \emph{Network} method, which starts from central nodes with high betweenness centrality, and the \emph{Content} method, which starts from users who have previously answered similar questions. Both methods use \emph{Random Walks} (RW) for exploration, beginning from selected \emph{seed nodes}. Through RW, TUEF selects a subset of the most relevant experts based on the question topic.
The results from these explorations are combined to create a set of potential experts. TUEF then extracts static features (such as user Reputation) and dynamic features (derived from the graph exploration) for each identified expert. Finally, it applies a decision tree-based Learning-to-Rank method to score and rank these candidates based on their expected relevance to the new question. For further details on the framework, we refer readers to \cite{amendola2024towards}.
This modular component-based approach allows us to analyze in isolation the possible gender bias originated by each information source, i.e., textual content, feature-based, network-based, discussed in Section \ref{sec:relatedEF}. 

\subsection{Analysis of the various steps of TUEF}

\begin{table}[]
\caption{Characteristics of the TUEF experiments.}
\centering
\resizebox{\columnwidth}{!}{
\begin{tabular}{cccccc}
\hline
Years & \#Tags & \#Layers & \begin{tabular}[c]{@{}c@{}}MinAccepted\\ Answers\end{tabular} & Experts & \%Females \\ \hline
2017-2018 & 23,583 & 8 & 22 & 2,335 & 8.39 \\
2019-2020 & 22,374 & 8 & 21 & 1,967 & 7.27 \\
2021-2022 & 19,580 & 10 & 19 & 1,454 & 9.15 \\
2017-2019 & 26,547 & 8 & 24 & 2,911 & 8.31 \\
2020-2022 & 23,543 & 8 & 21 & 2,234 & 8.06 \\
2017-2022 & 33,016 & 9 & 25 & 4,506 & 7.79 \\\hline
\end{tabular}}

\label{tab:tuef_statistics}
\end{table}\vspace{-4pt}

In this section, we analyze TUEF and its components from a gender bias perspective, and we analyze the corrective effect of RW, uncovering the role of \emph{homophily} as a method to balance or mitigate potential biases. Precisely, we assess gender balance for the Content and Network methods by examining users selected as seed nodes and those chosen during graph exploration through Random Walk (RW). To ensure comprehensive results, we conduct experiments on SO considering pairs of years, groups of three years, and the entire selected dataset (Table \ref{tab:dataset}). Table \ref{tab:tuef_statistics} presents various statistics, including the number of Tags in the training set, the number of Layers in the MLG, the minimum number of accepted answers (MinAcceptedAnswers) required for users to be selected as experts (corresponding to the 95th percentile), and the number of Experts selected along with the corresponding percentage of women (\%Females).

\paragraph{\textbf{Components Biases}}
\begin{figure}
    \centering
    \includegraphics[width=0.8\linewidth]{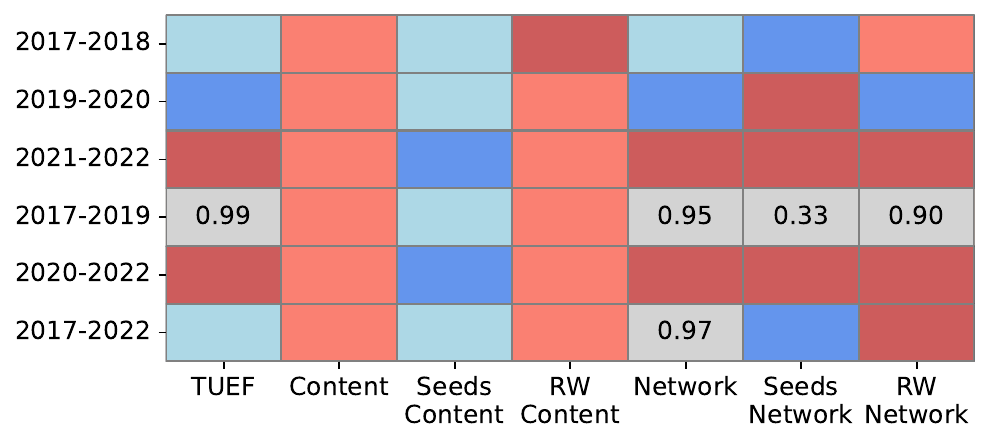}
    \caption{Visualization of gender overrepresentation across datasets (y-axis) and TUEF model components (x-axis)  using a t-test with \emph{greater} alternative for both genders. Cells show p-values only if they exceed \(\alpha=0.01\), indicating gender balance. Light red cells indicate women overrepresented by \(<0.5\%\), dark red by \(\le0.5\%\). Light blue and blue indicate the same for men, respectively.\vspace{-4pt}}\vspace{-4pt}
    \label{fig:overrep}\vspace{-4pt}
\end{figure}

Figure \ref{fig:overrep} illustrates gender overrepresentation across different TUEF components. The y-labels represent the experiments, while the x-labels denote the components: TUEF (entire framework), Content, and Network (both divided into \emph{Seeds} nodes and \emph{RW} nodes). The aim is to demonstrate the presence, or lack, of a statistically significant gender imbalance. To this end, we ran TUEF on 5000 queries. For each query, we computed the gender percentages and performed a t-test. Specifically, the null hypothesis is that the computed mean equals the population's average. Instead, the alternative hypothesis is that the computed mean is higher. If the p-value associated with the given alternative is lower than $\alpha$, the test rejects the null hypothesis in favour of the alternative, thus revealing a gender overrepresentation in this context. We set alpha=0.01 and performed 100 permutations for more accurate results.

In the figure, the cells contain the p-values only if they exceed $\alpha$ for both genders, thus demonstrating a gender balance. The other tests, instead, reject the null hypothesis in favour of the alternative, highlighting significant differences between the actual percentage of a specific gender and the expected one. Cells coloured light red indicate women are overrepresented with a difference of less than 0.5\% with respect to the expected average, while dark red indicates a difference higher than 0.5\%. Light blue and blue cells indicate male overrepresentation by less than and more than 0.5\%, respectively. 

The results show that the Seed Content method predominantly overrepresents men, which is evident in blue cells' prevalence. This suggests a bias in the initial seed node selection towards male nodes, which skews the representation. This result aligns with the statistics shown in Table \ref{tab:features_statistics}, where men provide more accepted answers, thus leading to more men being included in the Content results. In contrast, the RW Content method consistently overrepresents women, as indicated by the frequent red and light red cells. This implies that the RW algorithm favours female nodes once the initial selection is made, potentially due to connectivity patterns and edge weights leading to more female nodes being selected. The Network-based methods, both Seeds and RW, do not exhibit a consistent pattern in gender representation. These methods show mixed results, with instances of both male and female overrepresentation, suggesting that different factors influence them and may lead to a more balanced or variable representation.

The combination of the Content method, which is slightly biased against men, and the Network method, which does not show a systematic bias against any gender, allows TUEF to avoid being systematically biased against a specific gender. Additionally, cases with slight over-representation (difference less than 0.5\%) or gender balance occur more frequently than strong overrepresentation. The final gender balance or imbalance depends on the data considered. The results shown in Figure \ref{fig:overrep} and Table \ref{tab:features_statistics} are aligned, with TUEF overrepresenting women in the datasets where men and women have similar participation rates. 

Overall, the analysis reveals that while the seed nodes selected by the Content method may introduce initial biases, the RW algorithm can significantly alter representation dynamics.

\paragraph{\textbf{Biases in Random Walk}}

\begin{figure}
    \centering
    \includegraphics[width=0.8\linewidth]{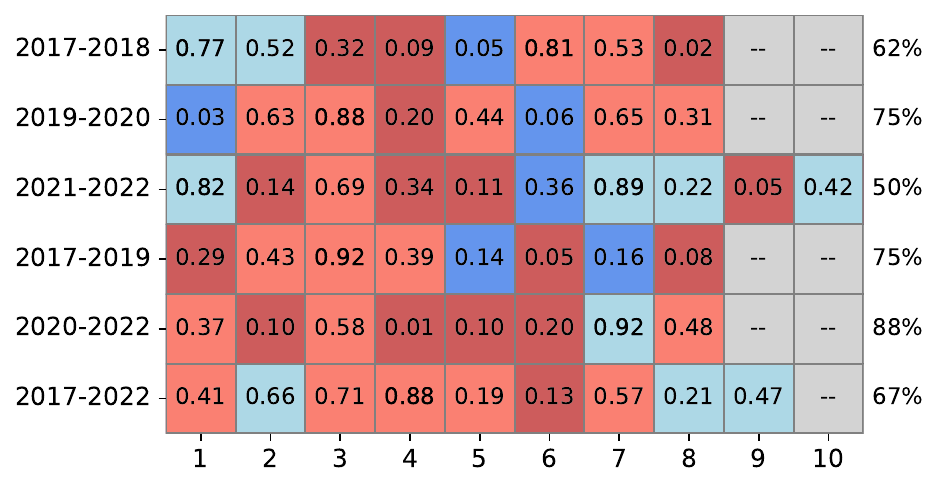}
    \caption{Welch's t-test ($\alpha=0.01$) comparison of average node weights between genders across datasets (y-axis) and MLG layers (x-axis). Red and light red cells indicate women's weights are higher by $\le1$\% and $<1$\%, respectively. Blue and light blue cells indicate the same for men. Colored cells show significant differences with p-values. Percentages on the left show layers where women's weights are higher.}\vspace{-12pt}
    \label{fig:avg_nodes_weights}
\end{figure}

Several studies have shown that women tend to create homophilic networks to create a supportive environment, boosting participation. Specifically, research by \cite{ford2017someone, morgan2017programming, brooke2021trouble} on SO demonstrated that women are more likely to respond to a question if they see responses from other women (peer-parity). Joining these results with the TUEF approach that models user interactions by considering their similarity in response behaviour under specific topics, we study why TUEF shifts from men overrepresentation to women overrepresentation due to the RW component. Given that RW is weighted, choosing the edge with the highest weight with the highest probability, is essential to examine the distribution of average weights for men and women nodes. The average weight of a node is computed as the sum of the weights of its edges divided by the number of edges.

The y-axis in Figure \ref{fig:avg_nodes_weights} represents the datasets, while the x-axis represents the number of layers, ranging from 1 to 10. For each dataset and each layer, we applied Welch's t-test with $\alpha=0.01$ and 1000 permutations to compare the average node weights for men and women. This test helps to determine if there are statistically significant differences between genders. The cells in the plot are colour-coded based on the statistical significance and the magnitude of the difference: red cells indicate that the average node weights for women nodes are at least 1\% higher than for men, while light red cells indicate a positive difference lower than 1\%. Blue and light blue cells model the same for men, respectively. These coloured cells indicate significant differences, and they report the corresponding p-values. Grey, empty cells represent cases where the graph does not have the whole 10 layers. Additionally, on the left side of each row, we report the percentage of layers where women have a higher average weight than men, regardless of the statistical differences.

The figure demonstrates the prevalence of red cells, indicating that women generally have higher average node weights than men.
This suggests that, once initial selections are made, the RW algorithm tends to favor female nodes, potentially due to connectivity patterns and edge weights leading to more female nodes being selected. The percentages on the left side of each row show that women have a higher average weight in more than 50\% of the layers, also exceeding 80\% in one dataset.

\begin{figure}
    \centering
    \includegraphics[width=0.8\linewidth]{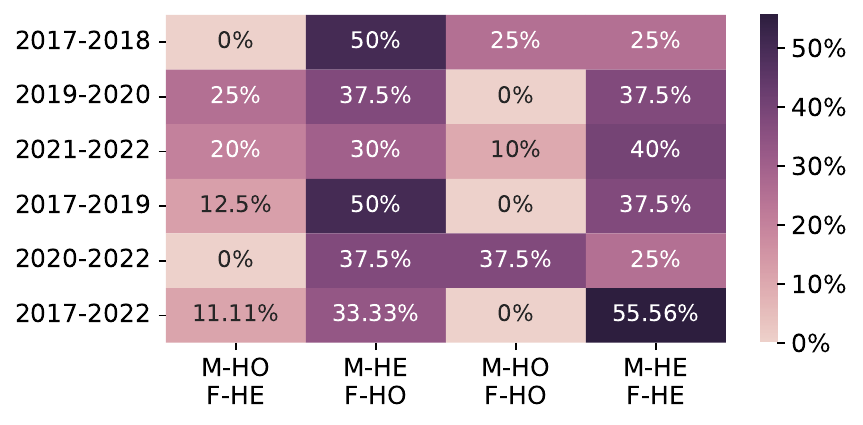}
    \caption{Heatmap for homophily analysis across all datasets (y-axis) and four scenarios (x-axis). M-HO stands for Male Homophilic, M-HE for Male Heterophilic, F-HO for Female Homophilic, and F-HE for Female Heterophilic. Each cell shows the percentages of layers where these scenarios occur.}
    \label{fig:homophily}
\end{figure}

The analysis depicted in Figure \ref{fig:homophily} highlights the dynamics of homophily in the TUEF MLG, revealing significant gender-based patterns in user connections. Using the definition of homophily formulated in \cite{fabbri2020effect}, we examine four scenarios: men being homophilic and women heterophilic (M-HO, F-HE); men heterophilic and women being homophilic  (M-HE, F-HO); both genders being homophilic (M-HO, F-HO); both genders being heterophilic (M-HE, F-HE). The figure shows the percentages of layers where these scenarios occur in each cell, with strong colours indicating higher percentages. Notably, the 'M-HE, F-HO' configuration shows significantly higher percentages with more intense colours compared to the 'M-HO, F-HE' scenario. This suggests that women tend to form homophilic networks more frequently than men, who are generally more heterophilic. The second most common scenario is that both genders are heterophilic. These findings explain why the RW method within TUEF shifts the representation balance in favour of women. Since men are more heterophilic and women have higher link weights on average, it is easy for the algorithm to transition from a male node to a female one. Once the RW algorithm begins favouring female nodes, it continues to select them due to their interconnectedness, reversing the initial male overrepresentation.

\paragraph{\textbf{Comparison between TUEF and TUEF}*}

\begin{figure}
    \centering
    \includegraphics[width=0.8\linewidth]{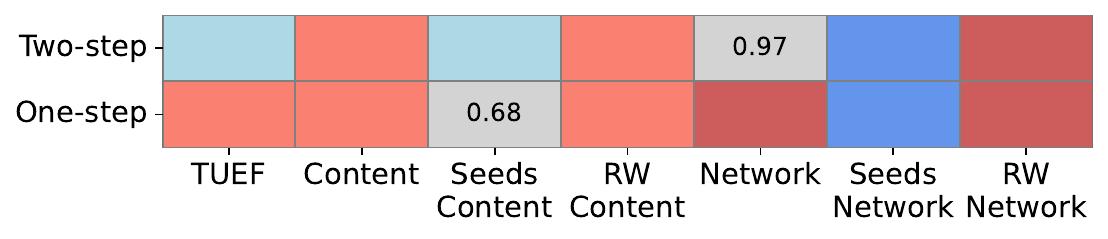}
    \caption{Visualization of gender overrepresentation across Two- and One-step preprocessig strategies and TUEF model components (x-axis) using a t-test with \emph{greater} alternative for both genders. Cells show p-values only if they exceed \(\alpha=0.01\), indicating gender balance. Light red cells indicate women overrepresented by \(<0.5\%\), dark red by \(\le0.5\%\). Light blue and blue indicate the same for men, respectively.\vspace{-4pt}}\vspace{-4pt}
    \label{fig:tuef_tuef*_overrep}
\end{figure}

\begin{figure}
    \centering
    \includegraphics[width=0.8\linewidth]{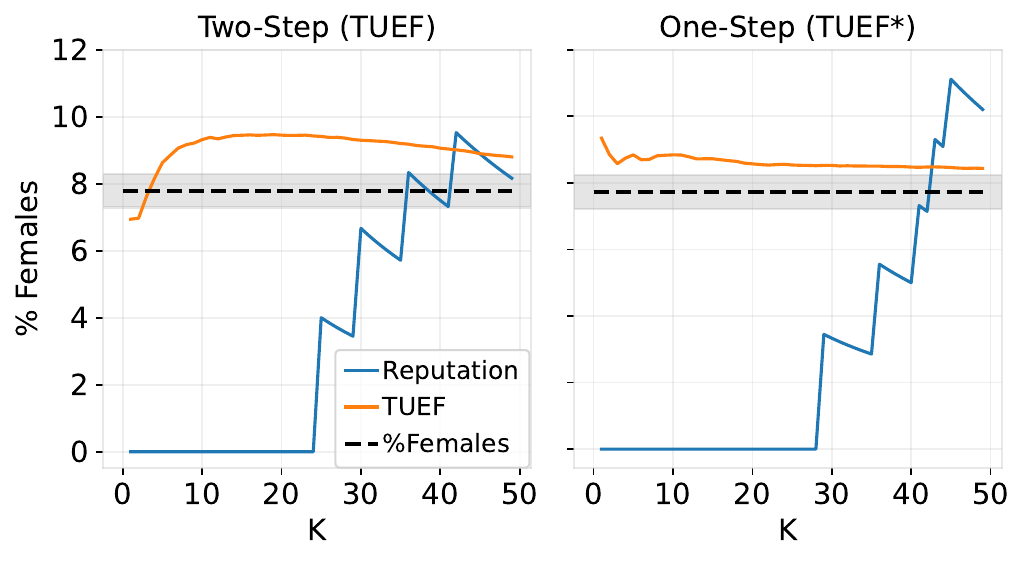}
    \caption{Comparison of TUEF (left) and TUEF* (right) on the entire dataset, showing percentages of women (y-axis) at different cut-off points in prediction lists (x-axis). The blue line represents users ordered by Reputation, and the black line indicates the percentage of women in the dataset.\vspace{-4pt}}\vspace{-4pt}
    \label{fig:tuef_tuef*_pred}
\end{figure}

To further verify the results found in section \ref{sec:preprocessing}, we compare TUEF and TUEF* on the entire dataset. TUEF automatically sets the minimum number of accepted answers an expert should have based on data. TUEF* considers the same number without further applying the acceptance ratio filter.
TUEF selected 4506 experts, while TUEF* 8403, both maintaining similar percentages of women (lower compared to the one reported in Table \ref{tab:dataset}). Figure \ref{fig:tuef_tuef*_overrep}, as Figure \ref{fig:overrep}, presents T-test results for gender overrepresentation. The two models appear very similar. However, TUEF ensures a balance when selecting seed nodes for the Content method but overrepresents women in the Network method.
The final result is the opposite: TUEF slightly overrepresents men, while TUEF* slightly overrepresents women.
Additionally, a predominance of women is observed in the nodes selected during RW, confirming the results in Figure \ref{fig:homophily}.
Figure \ref{fig:tuef_tuef*_pred} illustrates the percentages of women at different cut-off points in prediction lists for both models. The blue line represents users ordered by Reputation, with the black dashed line indicating the dataset's percentage of women. The orange line shows TUEF's predictions with Two- and One-step preprocessing.
Ordering by Reputation places women beyond the twentieth position, amplifying gender bias. However, TUEF*, by eliminating the acceptance ratio factor, aligns closer to the black line and stabilizes gender balance from $K=20$.
Moreover, Figure \ref{fig:tuef_tuef*_pred} suggests TUEF overrepresents women, while Figure \ref{fig:tuef_tuef*_overrep} indicates a slight overrepresentation of men. This discrepancy arises because Figure  \ref{fig:tuef_tuef*_pred} shows the top 50 users' ranking,  whereas TUEF slightly overrepresents men in the selected relevant experts subset.
Finally, the discrepancy between TUEF in Figure \ref{fig:perc_cutted_lists} (lower ranking for women) and Figure  \ref{fig:tuef_tuef*_pred} is due to TUEF's results depending on the data considered, rather than indicating a specific bias (as demonstrated also in Figure \ref{fig:overrep}).

\section{Conclusion}
The EF task is essential yet challenging for CQ\&A platforms like SO. Identifying the right expert users enhances user engagement and supports reputation-building. However, the current Reputation system in SO often favours men due to their higher activity levels. This bias does not necessarily reflect greater competence and can create barriers for minority groups, including women.
Our analysis shows that relying solely on Reputation metrics or UpVotes can exclude women. While various studies have tried incorporating content, relationships, and features for a broader perspective, minor adjustments can contribute to a fairer EF model.
Key findings include the impact of preprocessing strategies on gender representation. Among the methods, selecting the top 10\% of active users is the least balanced. Conversely, a one-step preprocessing approach with a minimum number of accepted answers improves gender balance. Including acceptance ratios tends to reduce women's representation due to men's generally higher activity levels.
Integrating both content and social information is vital. Relying solely on content tends to select more men, while social information might select more women due to peer-parity. Thus, to develop CQ\&A models that can balance accuracy and fairness, it is essential to understand the data and explicitly address biases in all the design phases.

\begin{acks}
To Robert, for the bagels and explaining CMYK and color spaces.
\end{acks}

\bibliographystyle{ACM-Reference-Format}
\bibliography{bib}

\end{document}